\documentclass[useAMS,usenatbib]{mn2e}
\usepackage{graphicx}\usepackage{epsfig}\usepackage{epsf}
\usepackage{amsmath}\usepackage{amssymb}\usepackage{stfloats}
\usepackage{caption} \usepackage{multirow} \usepackage{placeins}
\title{On the Common Envelope Efficiency}
\author[Z.-Y. Zuo and X.-D. Li]{Zhao-Yu Zuo$^{1,3}$\thanks{E-mail:zuozyu@gmail.com} and Xiang-Dong Li$^{2,3}$ \\
$^{1}$Department of Physics, School of Science, Xi'an Jiaotong University, Xi'an 710049, China\\
$^{2}$Department of Astronomy, Nanjing University, Nanjing 210093, China\\
$^{3}$Key laboratory of Modern Astronomy and Astrophysics (Nanjing University), Ministry of Education, Nanjing 210093, China\\
}

\begin{document}

\pagerange{\pageref{firstpage}--\pageref{lastpage}} \pubyear{2014}
\maketitle
\label{firstpage}

\begin{abstract}

In this work, we try to use the apparent luminosity versus displacement (i.e., $L_{\rm X}$ vs. $R$) correlation
of high mass X-ray binaries (HMXBs) to constrain the common envelope (CE) 
efficiency $\alpha_{\rm CE}$, which is a key parameter affecting the evolution of 
the binary orbit during the CE phase. The major updates that crucial for the CE 
evolution include a variable $\lambda$ parameter and a new CE criterion for 
Hertzsprung gap donor stars, both of which are recently developed. We find that, 
within the framework of the standard energy formula for CE and core 
definition at mass $X=10$\%, a high value of $\alpha_{\rm CE}$, i.e., around 
0.8-1.0, is more preferable, while $\alpha_{\rm CE}< \sim 0.4$ likely can not reconstruct 
the observed $L_{\rm X}$ vs. $R$ distribution. 
However due to an ambiguous definition for the core boundary in the literature, 
the used $\lambda$ here still carries almost two order of magnitude uncertainty, 
which may translate directly to the expected value of $\alpha_{\rm CE}$. We present 
the detailed components of current  HMXBs and their spatial offsets from star 
clusters, which may be further testified by future observations of HMXB populations 
in nearby star-forming galaxies.

\end{abstract}


\section{Introduction}

One of the most important stages in the evolution of close binary stars is
common envelope (CE) evolution. It is commonly thought to occur when
the expanding primary star transfers mass to its companion at a too high rate that
the companion cannot accrete it. This leads to the companion star being engulfed
by the envelope of the primary. The orbital energy and angular momentum of the orbiting
components are then transferred into the CE, resulting in the orbital decay and spiral-in
of the star \citep[for a recent review, see][and references therein]{Ivanova13}. Two main
schemes have been developed to describe the CE process. One (and the most popular one)
is $\alpha_{\rm CE}$-formalism \citep{Webbink84}. Usually the parameter $\alpha_{\rm CE}$
(see Eq~2 in this paper) in this scheme is introduced, to describe the efficiency of converting
the orbital energy into the kinetic energy of the envelope, resulting in the ejection of the
envelope. If there is sufficient energy ($\propto \alpha_{\rm CE}$) that can be used to
eject the envelope, the system will survive to be a compact binary, otherwise the two stars
coalesce instead. The other is  the so-called $\gamma$-algorithm, which considers angular
momentum transformation during the spiral-in \citep{nvyp00,nt05}. Despite extensive three
dimensional hydrodynamical simulations \citep[e.g.,][]{rl96,stc98,stb00,Fryxell00,OShea05,
Fryer06,Passy12,rt12}, the physics of the CE evolution still remains poorly understood. 

Many efforts have been made to constrain the value of
$\alpha_{\rm CE}$. An important way is to use individual post-common-envelope binaries
(PCEBs) to derive the magnitude of $\alpha_{\rm CE}$ \citep{maxted06}. Two
recent works in this respect were made by \citet{zgn00}
and \citet{marco11}, independently based on the Sloan Digital Sky
Survey (SDSS) data of white dwarf/main-sequence (WDMS) binaries. By reconstructing the
evolution of WDMS binaries, they can derive the value
of $\alpha_{\rm CE}$, i.e., around $0.2-0.3$. An important progress in their studies
is that they used a more realistic approximation for the envelope binding energy
(i.e., parameter $\lambda$, see Eq~2) than a constant one. Their results also imply that future
binary population synthesis (BPS) simulations with a proper treatment of
$\lambda$ is necessary and required urgently. An alternative way to constrain
$\alpha_{\rm CE}$ is to perform BPS simulations. By comparing the predictions with the
observed properties of PCEBs, the value of $\alpha_{\rm CE}$ could be properly
constrained. A lot of works have been carried out in this direction over the past two
decades. The research targets cover a large variety of different PCEB populations,
including WDMS \citep[][and see references therein]{pw06,pw07,dkw10},
extreme horizontal branch stars \citep{han02,han03,han08}, low-mass X-ray
binaries \citep{podsiadlowski03}, and high-mass X-ray binaries \citep[HMXBs,][]{zuo10,zuo13b}, etc.
However most of the previous BPS studies adopted a rather simplified
treatment of $\lambda$, as a constant, around 0.5 typically.
It is worth noting that among the studies mentioned above, \citet{dkw10} and \citet{zuo13b}
have used a more realistic treatment of $\lambda$ in their simulations. For example, \citet{zuo13b},
by incorporating variable values of $\lambda$ into the BPS code, constrained the CE mechanisms
based on the measured X-ray luminosity function of HMXBs, and found that the
$\alpha_{\rm CE}$-formalism is more preferable and the value of  $\alpha_{\rm CE}$
is likely high, i.e. $\sim 0.5-1$.

Note that, in our previous work \citep{zuo10}, indications already revealed
that the X-ray luminosity versus displacement (i.e., $L_{\rm X}$ vs. $R$)
distribution of HMXBs from star clusters in star-forming galaxies may provide a clue to
constrain the value of $\alpha_{\rm CE}$. Based on data from {\it Chandra\/}
and NICMOS on board {\it Hubble Space Telescope\/} ({\it HST\/}) of X-ray
sources and star clusters, respectively, \citet{kaaret04}
found that the X-ray sources are generally located near the star clusters, and
the brighter ones are preferentially closer to the clusters. Moreover there is an
absence of luminous sources ($L_{\rm X}>10^{38}$ erg\,s$^{-1}$) at
relatively large displacements ($>200$ pc) from the clusters. Based on this,
we have modeled the kinematic evolution of XRBs from the star clusters from a
theoretical point of view. It is shown that the parameter $\alpha_{\rm CE}$,
by affecting the binary orbit during the CE, can further determine the evolutionary
state of the donor star during the X-ray phase. Moreover the value of
$\alpha_{\rm CE}$ determines the orbital velocity and
the global velocity of the post-CE binary after the SN, hence the spatial offset of the system. So
different choices of $\alpha_{\rm CE}$ may give different $L_{\rm X}$ vs. $R$
distributions, as already illustrated in \citet{zuo10}. However the greatest limitation in the paper is
the oversimplified treatment of CE, especially the parameter $\lambda$, due to the
lack of an easy-to-use fitting formulae of $\lambda$ at that time, as discussed in the paper.

In the present work, we applied a most up-to-date BPS technique to revisit the
$L_{\rm X}$ vs. $R$ problem. The updates include several folds, with two crucial
for the CE phase (i.e., variable $\lambda$ and a new CE criterion for 
Hertzsprung gap (HG) donor stars, see \S 2.1.1).
The main goal of this study is to see whether the $L_{\rm X}$ vs. $R$
correlation could be reconstructed within the range of reasonable value of
$\alpha_{\rm CE}$, given the current knowledge of CE evolution. We
present the $L_{\rm X}$ vs. $R$ distribution under different choices of
$\alpha_{\rm CE}$, by comparing with observations, to constrain the value of
$\alpha_{\rm CE}$. The well simulated $L_{\rm X}$ vs. $R$
distribution can also help to understand the nature of the sources
and may be testified by future high-resolution X-ray and optical observations.

This paper is organized as follows. In \S 2 we describe the
BPS method and the input physics for X-ray binaries (XRBs) in our
model. The calculated results and discussions are presented in \S 3. Our conclusions
are in \S 4.

\section{MODEL DESCRIPTION}

\subsection{Assumptions and input parameters}

\subsubsection{binary evolution}

To follow the evolution of HMXBs, we have used the BPS code initially developed
by \citet{Hurley00,Hurley02} and recently updated by \citet{zuo13} in our calculation.
In the present code, the compact object masses are calculated using a prescription same as in
\citet[][i.e., the rapid supernova mechanism]{fryer12}, which can model the mass distribution of
neutron stars (NSs) and black holes (BHs) in the Galactic XRBs quite well \citep{bel12}.
We also consider the formation of NSs through
electron capture supernovae \citep[i.e. ECS,][]{Podsiadlowski04}. The maximum NS mass is assumed to be $2.5\,M_{\odot}$,
above which black hole (BH) is assumed to form. We account for fallback and direct BH formation
during core collapse \citep{fk01}. The mass loss prescription for the winds in massive stars
has also been updated \citep[][also see Belczynski et al. 2010]{vink01}.

Following \citet{zuo10}, we adopt a constant star formation rate for a duration of 20 Myr,
in parallel with \citet{kaaret04}, to model the HMXB populations. We construct several
models, in each model, we evolve $8\times10^6$ primordial systems, all of which are
initially binary systems. We set up the same grid of the initial parameters (i.e., the primary
mass, secondary mass and orbital separation) as \citet{Hurley02} and evolve each binary
then. In the following we describe the assumptions and input parameters in our basic model
(i.e., model A09 in Table 1, equivalent to model M1 in Table 2).

\noindent {\em (1) initial parameters}\\
We adopt the initial mass function (IMF) of \citet[][hereafter KROUPA01 for short]{Kroupa01} for the
distribution of the primary mass ($M_1$). For the secondary's mass
($M_2$), a power law distribution of the mass ratio $P(q)\propto q^{\alpha}$ is assumed,
where $q\equiv M_2/M_1$. In our basic model, a flat distribution is assumed,
i.e. $\alpha=0$. Finally a uniform distribution is also taken for the logarithm of the orbital
separation (i.e., $\ln a$).

\noindent {\em (2) CE evolution} \\
The stability of mass transfer in binaries depends on both the mass ratio and
the evolutionary states of both stars. Often a critical ratio of the donor mass to
the accretor mass, $q_{\rm crit}$, is defined, above which mass transfer is dynamically
unstable between the two components. The ratio $q_{\rm crit}$ usually varies with the
evolutionary state of the donor star at the onset of RLOF \citep{hw87,w88,prp02,ch08}.
In this study, we adopt an updated $q_{\rm crit}$ (which was numerically calculated
for a large number of binaries by use of an updated version of the Eggleton (1971, 1972)'s
stellar evolution code) for CE initiated by HG donor
stars \citep[][also see the Appendix A in Zuo, Li \& Gu 2014a]{shao12}.
If the primordial primary is on the first giant branch (FGB) or the
asymptotic giant branch (AGB), we use
\begin{equation}
q_{\rm crit}=[1.67-x+2(\frac{M_{\rm c1}}{M_1})^5]/2.13
\end{equation}
where $M_{\rm c1}$ is the core mass of the donor star, and $x$=d\,ln$R_1/$d\,ln$M$ is
the mass-radius exponent of the donor star. If the donor star is a
naked helium giant, $q_{\rm crit}=0.784$ \citep[see][for more details]{Hurley02}.

Usually the CE interaction is parameterized in terms of the orbital energy and the binding energy.
It is expressed as $E_{\rm bind} \equiv \alpha_{\rm CE} \triangle E_{\rm orb}$ \citep{Webbink84,webbink08},
where the CE parameter $\alpha_{\rm CE}$ describes the efficiency of converting orbital energy
(\textbf{$E_{\rm orb}$}) into kinetic energy, to expel the CE, and $E_{\rm bind}$ is the binding energy of the
envelope. We adopt the standard energy prescription presented by
\citet{Kiel06} to compute the outcome of
the CE phase\footnote{Another formulation \citep{Webbink84} takes the initial orbital energy 
(i.e., the second term in parenthesis in Eq.~2) to be $-\frac{GM_{1}M_{2}}{2 a_{\rm i}}$, 
but gives negligible difference in the final results. }, 
\begin{equation}
\alpha_{\rm CE}\left(\frac{GM_{\rm c}M_{2}}{2 a_{\rm f}}-\frac{GM_{\rm
c}M_{2}}{2 a_{\rm i}}\right)=-\frac{GM_1M_{\rm env}}{R_{L_1}\lambda},
\end{equation}
which yields the ratio of final (post-CE) to initial (pre-CE) orbital
separations as
\begin{equation}
\frac{a_{\rm f}}{a_{\rm i}}= \frac{M_{\rm c}M_{2}}{M_{1}}
\frac{1}{M_{\rm c}M_2/M_1+2M_{\rm env}/(\alpha_{\rm CE} \lambda
R_{\rm L1})}.
\end{equation}
where $G$ is the gravitational constant, $M_{\rm c}$ is the helium-core mass of the
primary star, $R_{L_1}$ is the RL radius of the primary star, $M_{\rm env}$ is
the mass of the primary's envelope, $a_f$ and $a_i$ denote the final and initial orbital
separations, respectively.

Here $\lambda$ depends on the structure of the primary star, conventionally
chosen as constant \citep[typically $\sim 0.5$,][]{Hurley02,zuo10} but in reality
possibly far from it \citep{dewi00,sluys06}. We use the fitting formulae of 
envelope binding energy $E_{\rm bind}$ presented by \citet{loveridge11} 
which implicitly include variable $\lambda$. The $E_{\rm bind}$ is calculated 
by integrating the gravitational and internal energies from the core-envelope 
boundary to the surface of the star $M_{\rm s}$ as follows,
\begin{equation}
E_{\rm bind}=\int^{M_s}_{M_c}(-\frac{Gm}{r(m)})dm+\alpha_{\rm in}\int^{M_s}_{M_c}E_{in}dm,
\end{equation}
where $E_{in}$ is the internal energy per unit of mass, containing the thermal energy of
the gas and the radiation energy, but not the recombination energy \citep[for details, see][]{sluys06}.
Here we define $\alpha_{\rm in}$ the percentage of the internal energy contributing to
the ejection of the envelope. We take its value as 1 in our basic model and change it
to zero (models M6 and M13, see Table~2) for comparisons.

Note that the value of $\lambda$ is still uncertain due to an 
ambiguous definition of the stellar core boundary in the literature. Here the core is 
defined as the central mass below the location where $X=10$\% 
\citep{dewi00}. However other definitions are still possible to decide  
the boundary between the core and the envelope, for example, based on 
the effective polytropic index profiles \citep{hw87}, the change in 
the mass-density gradient \citep{Bisscheroux88}, the method suggested 
by \citet{han94}, and the entropy profile \citep{td01}, etc.
Different definitions of the core may give different values of 
$\lambda$, then translate directly similar uncertainties in the value of 
$\alpha_{\rm CE}$. In extreme cases, the `entropy profile' definition for the 
core can increase the value of $\lambda$ by a factor of 10-70 for massive stars 
\citep[][see also Ivanova 2011]{td01}, which may be the progenitor of 
HMXBs, then the expected value of $\alpha_{\rm CE}$ will be extremely small.

\begin{table*}\begin{center}
\caption{Different models on the treatment of the CE parameters $\alpha_{\rm CE}$. } 
\renewcommand\tabcolsep{5pt}
\begin{tabular}[b]{@{}cccccccccccc}\hline\hline
   Model                     & A02  & A03  & A04  & A05  & A06  & A07  & A08  & A09  & A10  \\  \hline
$\alpha_{\rm CE}$ & 0.2 & 0.3 & 0.4 & 0.5& 0.6 & 0.7 & 0.8 & 0.9 & 1.0 \\      \hline
\end{tabular}\label{tab:m9} 
\end{center}
\end{table*}

\begin{table*}\begin{center}
\caption{Parameters adopted for each model. Here $\alpha_{\rm CE}$
is the CE parameter, $q$ is the initial mass ratio, $\sigma_{\rm
kick}$ is the dispersion of kick velocity, IMF is the initial mass function, $\alpha_{\rm in}$
represents the percentage of the internal energy contributing to the ejection of the envelope,
STDw is the standard stellar winds, while STGw represents the standard wind mass-loss
rate increased to 200 per cent, STDm is the standard mass transfer rate (MTR) before the CE,
while REDm represents the standard MTR reduced to 50 per cent.} 
\renewcommand\tabcolsep{5pt}
\begin{tabular}{cccccccc}\hline\hline
   Model & $\alpha_{\rm CE}$ & P(q)    & $\sigma_{\rm kick}$ & IMF  & $\alpha_{\rm in}$ & winds  &  MTR \\
          &              &   & km/s & \\ \hline
      M1 & 0.9 & $\propto q^{0}$ & 150      &   KROUPA01    &  1   &  STDw   & STDm\\
      M2 & 0.9 & $\propto q^{0}$ & 150      &   MT87              &  1   &  STDw   & STDm\\
      M3 & 0.9 & $\propto q^{1}$ & 150      &   KROUPA01    &  1   &  STDw   & STDm\\
      M4 & 0.9 & $\propto q^{0}$ & 265      &   KROUPA01    &  1   &  STDw   & STDm\\
      M5 & 0.9 & $\propto q^{0}$ & 150      &   KROUPA01    &  1   &  STDw   & REDm\\
      M6 & 0.9 & $\propto q^{0}$ & 150      &   KROUPA01    &  0   &  STDw    & STDm\\
      M7 & 0.9 & $\propto q^{0}$ & 150      &   KROUPA01    &  1   &  STGw   & STDm\\   \hline
      M8 & 0.2 & $\propto q^{0}$ & 150      &   KROUPA01    &  1   &  STDw   & STDm\\
      M9 & 0.2 & $\propto q^{0}$ & 150      &   MT87              &  1   &  STDw   & STDm\\
      M10 & 0.2 & $\propto q^{1}$ & 150      &   KROUPA01    &  1   &  STDw   & STDm\\
      M11 & 0.2 & $\propto q^{0}$ & 265      &   KROUPA01    &  1   &  STDw   & STDm\\
      M12 & 0.2 & $\propto q^{0}$ & 150      &   KROUPA01    &  1   &  STDw   & REDm\\
      M13 & 0.2 & $\propto q^{0}$ & 150      &   KROUPA01    &  0   &  STDw    & STDm\\
      M14 & 0.2 & $\propto q^{0}$ & 150      &   KROUPA01    &  1   &  STGw   & STDm\\    \hline
\end{tabular}\label{tab:m14} 
\end{center}
\end{table*}

We consider several different choices of $\alpha_{\rm CE}$ (see Table~1) in order to constrain the value
in our calculations. The two extreme values of $\alpha_{\rm CE}$ are adopted as
0.2 and 1.0, since firstly $\alpha_{\rm CE}$ should be within unity  as we have considered the
potential internal energies in the quantity of $E_{\rm bind}$, and secondly the value of $\alpha_{\rm CE}< \sim 0.1$
is likely excluded according to our HMXB XLF modeling \citep{zuo13b}. These models are denoted as A02-A10,
respectively, where the last two digits correspond to the value of $\alpha_{\rm CE}$, and we set A09 as our basic model
according to our calculations.

\noindent {\em (3) SN kicks}\\
The newborn NS/BH in HMXBs may receive different
velocity kicks. For NS systems, the kick velocity $v_{\rm k}$ is assumed to
follow a Maxwellian distribution
\begin{equation}
   P(v_{\rm k})=\sqrt{\frac{2}{\pi}}\frac{v^{2}_{\rm k}}{\sigma_{\rm kick}^{3}}
   \exp(-\frac{v^{2}_{\rm k}}{2\sigma_{\rm kick}^{2}}),
\end{equation}
and we adopt $\sigma_{\rm kick}=150\rm\, km\,s^{-1}$ in our basic model though its value 
is still very uncertain. For BH systems, the natal kicks are assumed to multiplied by a factor of (1-$f_{\rm b}$)
if formed with partial mass fallback,  where $f_{\rm b}$ is the fraction of the stellar
envelope that falls back after the SN explosion. Specially BHs formed with small amounts of fall
back ($M_{\rm fb}<0.2\,M_{\odot}$) are assumed to receive full kicks. In situations where BHs
form silently (without a SN explosion) via direct collapse, no natal kick is adopted \citep{fryer12,Dominik12}.
Additionally, for ECS NSs, no kick velocity is assumed since these
are weak SN occurring for the lowest stars \citep[$M_{\rm ZAMS}=7.6 - 8.3 M_{\odot}$,][]{Hurley00,et04a,et04b,bel08}.

The velocity ($\textbf{v}_{\rm sys}$) of the binary system after the SN is determined by both the
natal kick and the orbital velocity of the system. It can be expressed as follows \citep[see][for details]{Hurley02},
\begin{equation}
   \textbf{v}_{\rm sys}=\frac{M_1^{'}}{M_{\rm b}^{'}}\textbf{v}_{\rm
   k}-\frac{\bigtriangleup M_1M_2}{M_{\rm b}^{'}M_{\rm b}}\textbf{v},
\end{equation}
where $M_1^{'}=M_1-\Delta M_1$ is the current mass of the primary star after losing mass $\Delta M_1$
during the SN, $M_{\rm b}=M_1+M_2$ and $M_{\rm b}^{'}=M_{\rm b}-\Delta M_1$ are the total masses
of the system before and after the SN, respectively;   $\textbf{v}$ is the relative orbital velocity of the
stars (see Eq.~A1 in Hurley et al. 2002). Tidal effect is also taken into account to remove any
eccentricity induced in a post-SN binary prior to the onset of mass transfer.

Several key parameters may affect the $L_{\rm X}$ vs. $R$ distribution,
such as the IMFs of the primary and secondary stars, the natal kick velocity, etc.
\citep[see][for more details and references therein]{zuo10}. So we also
adopt a top flatter IMF of \citet[][hereafter MT87, models M2 and M9]{mt87},
a ``twins" model (i.e., $\alpha=1$, models M3 and M10) and a higher dispersion
of kick velocity $\sigma_{\rm kick}=265$ $\rm km\,s^{-1}$ \citep[i.e., models
M4 and M11,][]{Hobbs05} to test their effects. Two additional models
with $\alpha=-1$ ($0.1\leq q \leq 1.0$) and orbital distribution $(\ln a)^{-0.45}$
\citep{sana13} have also been examined, the results of which are similar to
that of our basic model. Additionally, stronger wind
mass loss may reduce the envelope mass sufficiently, avoiding the occurrence of CE or CE
mergers \citep{Soker04}, so we adopt a wind mass loss rate enhanced
by a factor of two to test the effect (models M7 and M14). During the mass
transfer phase, mass and angular momentum loss rates may also affect the
following CE, so we assume half of the transferred mass lost in models M5 and
M12 (see Table~2).

\subsubsection{binary motion}
Since a star cluster in a star-burst region is usually centrally
concentrated, we assume a spherical potential and use the
cylindrical coordinate system ($r$, $\phi$, $z$) centered at the
cluster's center. The potential of a cluster can be described
as
\begin{equation}
   \Phi(r,z)=\frac{-GM}{\sqrt{r^2+z^2}+h},
\end{equation}
where $h$ is the half light radius and $M$ is the total mass of stars within $h$.
For typical star clusters we adopt $M=1.0\times10^{6} M_{\odot}$, and $h=3$ pc
\citep{ho96a,ho96b} in our calculations\footnote{We also reduced the cluster
mass to 50\%, and found no significant difference in
the final results.}. Then all stars are assumed to born uniformly in
the star cluster. The direction of its initial velocity is randomly designated,
which gives the initial velocity vectors $v_{\rm r}$, $v_{\phi}$, $v_{\rm z}$.
Because of the cylindrical symmetry of cluster potential, two space coordinates
$r$ and $z$ are sufficient to describe the HMXB distributions. We then integrate the motion
equations \citep[see Equations 19a and 19b in][]{paczynski90} with a
fourth-order Runge-Kutta method to calculate the trajectories of
the binary systems and collect the parameters of current XRBs if
turning on X-rays. Finally the positions of XRBs are projected  on the
$\phi=0$ plane to get the projected distances from star
clusters, i.e., $R=((r \cos \varphi)^2+z^2)^{1/2}$ with $\varphi$
uniformly distributed between 0 and $2\pi$. In our
calculations, the accuracy of integral is set to be $10^{-6}$ and
controlled by the energy integral.

\subsection{X-ray luminosity and source type}
We use the same methods to compute the $0.5-8$ keV X-ray
luminosity of different HMXB populations as in \citet{zuo13}.
In this study, we do not consider low-mass X-ray binaries
(LMXBs, donor mass $<2\,M_{\odot}$) since it takes $\sim$
Gyr for them to form, much longer than the duration time we considered.
Every accreting HMXB is usually powered by either accretion
disk or stellar winds. We use the classical \citet{Bondi44}'s formula to calculate
the mass transfer rate for wind-fed systems. In the case of disk accretion,
the material is transferred to the compact star by Roche-lobe overflow (RLOF).
We discriminate persistent and transient sources using the criteria of
\citet[][i.e., Eq~36 therein]{l01} for MS and red giant stars in RLOF cases.
The simulated X-ray luminosity is described as follows:
\begin{eqnarray}
&&L_{\rm X, 0.5-8 keV}\nonumber\\
&&=\left\{
\begin{array} { ll}
  \eta_{\rm bol}\eta_{\rm out}L_{\rm Edd}& \rm transients\ in\ outbursts, \\
  \eta_{\rm bol}\min(L_{\rm bol},\eta_{\rm Edd}L_{\rm Edd})& \rm persistent\ systems,
\end{array}
\right.
\end{eqnarray}
where $L_{\rm bol} \simeq 0.1\dot{M}_{\rm acc}c^2$ where $\dot{M}_{\rm acc}$
is the average mass accretion rate and $c$ the speed of light;
$\eta_{\rm bol}$ is the bolometric correction factor which
converts the bolometric luminosity ($L_{\rm bol}$) to the $0.5-8$ keV
X-ray luminosity, adopted as 0.2 for NS-XRBs and 0.4 for BH-XRBs, respectively  
though its range is $\sim 0.1-0.8$ for different types of XRBs \citep{bel08}; 
the critical Eddington luminosity $L_{\rm Edd} \simeq 4\pi GM_{1}m_{\rm
p}c/\sigma_{T}=1.3 \times 10^{38}m_{1}$\,erg\,s$^{-1}$ (where
$\sigma_{T}$ the Thomson cross section, $m_{\rm p}$ the proton
mass, and $m_{1}$ the accretor mass in the units of solar mass). $\eta_{\rm Edd}$ is called as
the `Begelman' factor which examines the allowed maximum super-Eddington accretion rate.
Here we adopt $\eta_{\rm Edd, NS}=5$ for NS XRBs
and $\eta_{\rm Edd, BH}=100$ for BH XRBs in our calculation \citep{zuo13}.
For transient sources the outburst luminosity is taken as a fraction ($\eta_{\rm out}$) of the critical
Eddington luminosity. We take $\eta_{\rm out}=0.1$ and 1 for NS and BH transients with
orbital period $P_{\rm orb}$ less and longer than 1 day and 10 hr, respectively \citep{chen97,Garcia03,bel08}.

The Be/X-ray binary (Be-XRB) is also included in our calculation. It 
contains a Be companion, usually accreted by an orbiting NS at its 
periastron, showing as X-ray transients. Five criterions are set to define 
it in a phenomenological way as in \citet[][also see Belczynski \& Ziolkowski, 2009]{zuo13}.
Technically, we randomly selected only 25\% \citep[$f_{\rm Be}=0.25$,][]{s88,z02,mg05} of NS
binaries hosting a massive ($3.0\,M_{\odot}-20.0\,M_{\odot}$) B/O star to predict their
numbers. The X-ray luminosity of a Be-XRB is calculated using Eq.~11 in
\citet{dll06}, which is based on data compiled by \citet{rp05}. Due to
the transient characteristics \citep[$\sim 0.2-0.3 P_{\rm orb}$,][]{reig11},
we adopt $DC_{\rm max}=0.3$ to give the maximum expected numbers.

\section{Results}

Based on a population of 66 X-ray point sources, \citet{kaaret04} studied the spatial offsets
between these sources and the star clusters in three starburst galaxies (i.e., M82, NGC 1569 
and NGC 5253). They found that the X-ray sources are preferentially located near the star 
clusters, with the brighter sources closer to the clusters. Moreover, they found no luminous 
source ($L_{\rm X}>10^{38}$ ergs$^{-1}$) at relatively large displacements ($>200$ pc) from 
the clusters. Here we modeled the kinematic evolution of XRBs in clusters, and presented the 
results below.

As stated before several models are constructed to investigate how the final results are 
impacted by the CE parameter $\alpha_{\rm CE}$. In practice, the following 
steps are performed to make the comparison. Step (I), the average source displacement 
distributions are calculated using the observed data. We restrict the source luminosities 
to the range of $10^{36}-10^{38}\rm erg\,s^{-1}$ in order to compare with \citet{kaaret04}. 
Step (II), a two-dimensional Kolmogorov-Smirnov (2D K-S) test is performed, to compare
the simulated displacement distributions with the average displacement derived in step I. 
This way, we can get a preliminary decision for the parameter $\alpha_{\rm CE}$, by 
seeing the derived possibility (i.e., $p$-value, see Tables~3 and 4. The value of $p$ that is 
less than $\sim 0.01$ suggests a significant different distribution.). Step (III), the $L_{\rm X}$ vs. $R$ 
distribution gives the occurrence possibilities of sources in each region, direct 2D K-S test with discrete 
sources is likely impossible, so we use it in a phenomenological way to further discriminate 
between models (i.e., the secondary check for the models).


\begin{table*}
\caption{Two-dimensional K-S test for models A02-A10. } \centering
\begin{tabular}{cccccccccc}\hline\hline
   Model                     & A02  & A03  & A04    & A05  & A06  & A07  & A08  & A09  & A10  \\   \hline 
   $p$-value            & $3\times 10^{-7}$ & $8\times 10^{-7}$ & $8\times 10^{-6}$   & 0.05 & 0.18 & 0.02 & 0.31 & 0.45 & 0.43 \\   \hline 
     \end{tabular}\label{tab:ks9}
\end{table*}

\begin{figure*}
  \centering
   \includegraphics[width=4.5in]{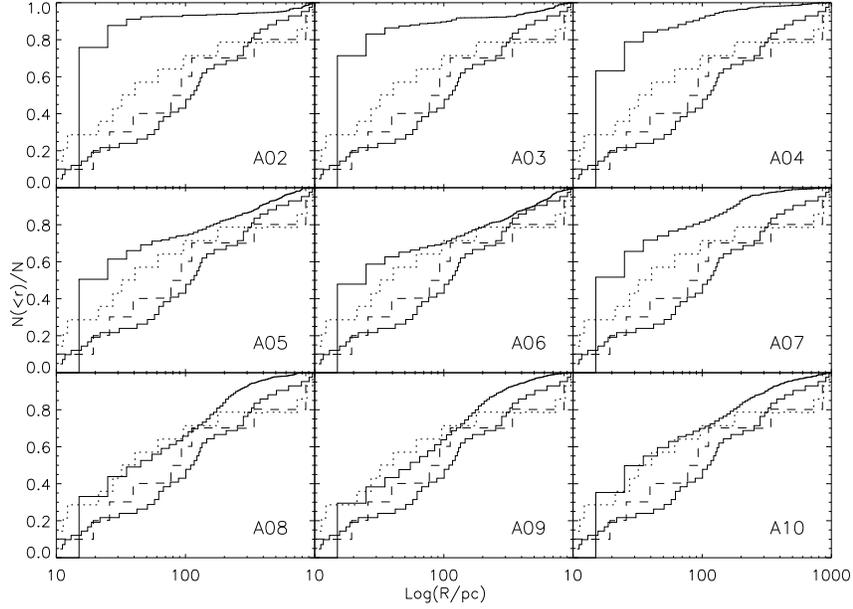}
\caption{The normalized cumulative distribution (thick-solid line) for models A02-A10, respectively (see Table~1).
The thin-solid, dotted and dashed lines represent the observed cumulative distributions of source displacements
in galaxies M82, NGC 1569 and NGC 5253 \citep[see Fig.~2 in][]{kaaret04}, respectively.}
  \label{Fig. 1}
\end{figure*}

\begin{figure*}
  \centering
   \includegraphics[width=4.5in]{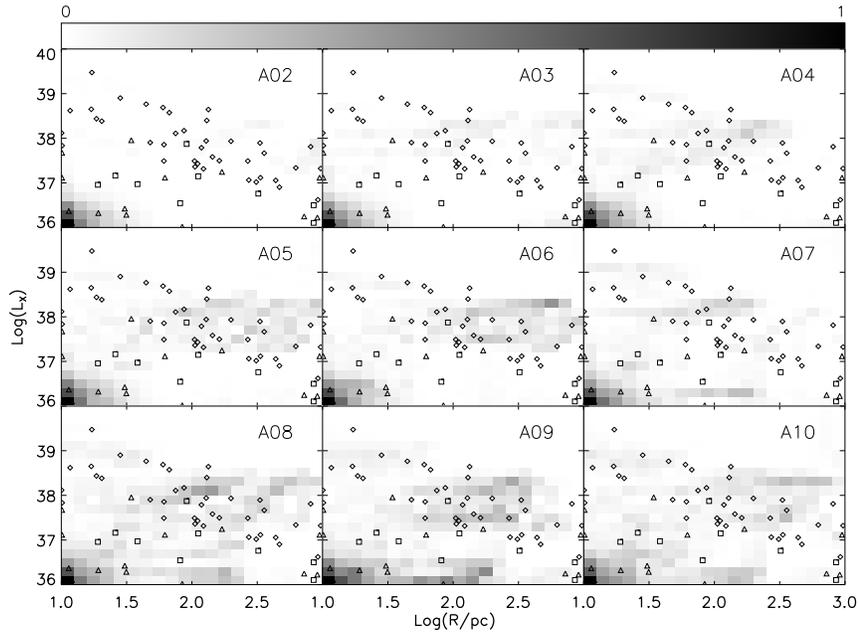}
\caption{The $L_{\rm X}-R$ distribution for models A02-A10 (see Table~1), respectively. Sources from M82,
NGC 1569 and NGC 5253 in \citet[][see Fig.~3]{kaaret04} samples are also shown as diamonds, triangles
and squares, respectively for comparison. }
  \label{Fig. 2}
\end{figure*}

Fig.~1 shows the simulated cumulative distribution of the XRB displacements at the age 
of 20 Myr for models A02-A10, respectively. We normalized each histogram by the total 
number of HMXBs in the 10-1000 pc region. Our results 
show that models with $\alpha_{\rm CE}$ between $\sim 0.8-1.0$ can match the 
observation \citep[e.g., Fig.~2 in][]{kaaret04} quite closely, while models with 
$\alpha_{\rm CE}<\sim 0.4$ (possibilities $<10^{-6}$, see Table~3) clearly fail. 
It shows that very few sources can move to 100 pc away from the star clusters, which is
in marked contrast with the observations. Models A05-A07, with possibilities $\sim 10^{-1}$, 
can not be firmly ruled out. Further model-check comes from Fig.~2 for the simulated 
distributions of X-ray luminosities ($L_{\rm X}$) at different displacements ($R$) 
from the star cluster in models A02-A10, respectively. The color bar represents the normalized
number ratio of HMXBs in the $L_{\rm X}-R$ plane. Note that the predicted $L_{\rm X}$
vs. $R$ correlation in models A08-A10 are compatible with the observations quite well, while
others likely fail, especially for models A02-A04, the $L_{\rm X}$ vs. $R$ correlation can 
hardly be reconstructed.  We find that this result is also consistent with the one 
obtained through HMXB XLF simulations recently presented by \citet{zuo13b}. 
Recently \citet{Ivanova11b} suggested a modified approach to the standard 
energy CE model, i.e., `enthalpy' prescription. A non-negative energy (i.e., $P/\rho$) 
is considered additionally to unbind the envelope. This may mean a decrease of the 
binding energy of the envelope, and hence the required CE efficiency $\alpha_{\rm CE}$, 
or an increase of the portion of the orbital energy used to eject the CE, equivalent 
to increase the effective value of $\alpha_{\rm CE}$. Therefore the range of 
$\alpha_{\rm CE}$ we obtained here still suffers from uncertainties. According to 
\citet{Ivanova11b}, the variation ratios in most giants are from $\sim 2$ to $\sim 5$.

\begin{table*}\begin{center}
\caption{Two-dimensional K-S test for models M1-M14. } 
\renewcommand\tabcolsep{5pt}
\begin{tabular}{cccccccc}\hline\hline
   Model                     & M1  & M2  & M3   & M4  & M5  & M6 &  M7  \\  
   $p$-value                          & 0.45 & 0.07 & 0.04 & 0.9 & 0.8 & 0.2   & $2\times 10^{-4}$ \\      \hline 
   Model                     & M8  & M9  & M10   & M11  & M12  & M13 &  M14  \\  
   $p$-value                          & $3\times 10^{-7}$ & $2\times 10^{-8}$ & $2\times 10^{-8}$ & $6\times 10^{-5}$ & $1\times 10^{-2}$ & $8\times 10^{-3}$   & $8\times 10^{-6}$ \\      \hline 
\end{tabular}\label{tab:ks14} 
\end{center}
\end{table*}

\begin{figure*}
  \centering
   \includegraphics[width=4.3in]{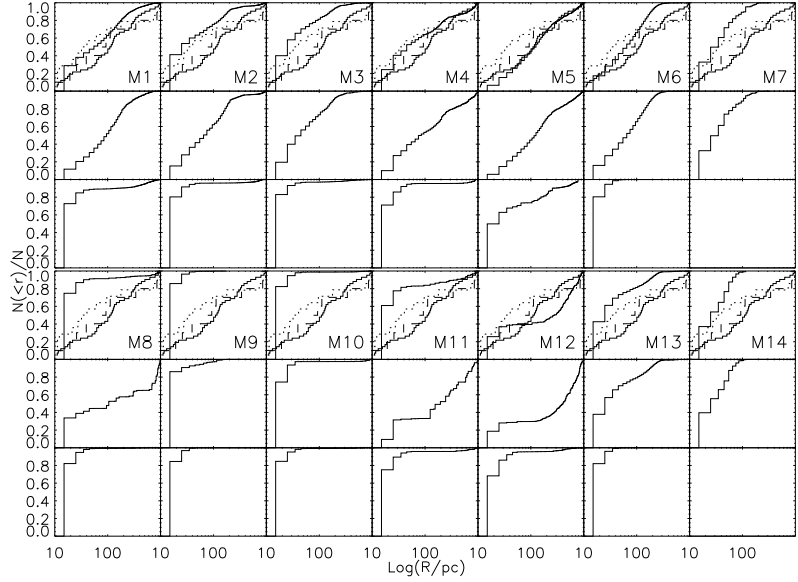}
\caption{The normalized cumulative distribution for the numbers of
ALL-XRBs (top), NS-XRBs (middle) and BH-XRBs (bottom) in models M1-M7 (upper panel)
and M8-M14 (lower panel), respectively. The thin-solid, dotted and dashed lines
represent the observed cumulative distributions of source displacements
in galaxies M82, NGC 1569 and NGC 5253 \citep[see Fig.~2 in][]{kaaret04}, respectively.}
  \label{Fig. 3}
\end{figure*}

\begin{figure*}
  \centering
   \includegraphics[width=4.3in]{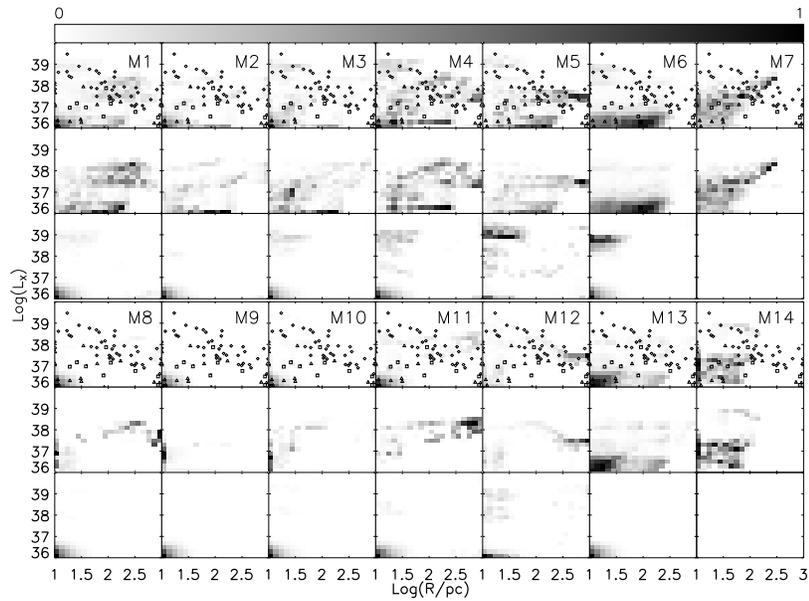}
\caption{The $L_{\rm X}-R$ distribution for ALL-XRBs (top), NS-XRBs
(middle) and BH-XRBs (bottom) in models M1-M7 (upper panel)
and M8-M14 (lower panel), respectively. Sources from M82,
NGC 1569 and NGC 5253 in \citet[][see their Fig.~3]{kaaret04} samples are also shown as diamonds, triangles
and squares, respectively for comparison.}
  \label{Fig. 4}
\end{figure*}

\begin{figure*}
  \centering
   \includegraphics[width=4.3in]{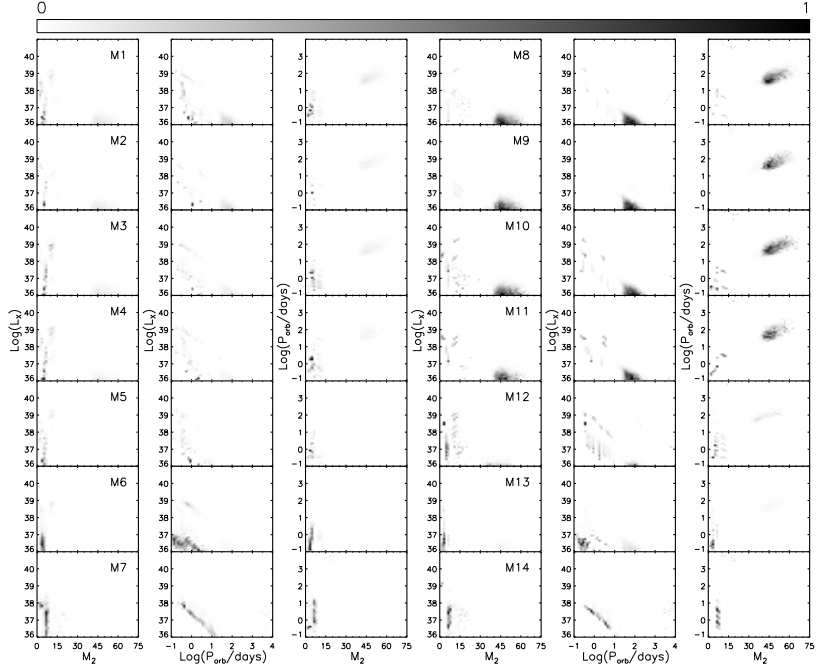}
\caption{The $L_{\rm X}-M_2$, $L_{\rm X}-P_{\rm orb}$, and $P_{\rm
orb}-M_2$ distributions in the $10<R<100$ pc region for models M1-M7 (left three columns,
from top to bottom) and models M8-M14 (right three columns, from top to bottom) ,
respectively.}
  \label{Fig. 5}
\end{figure*}

\begin{figure*}
  \centering
   \includegraphics[width=4.3in]{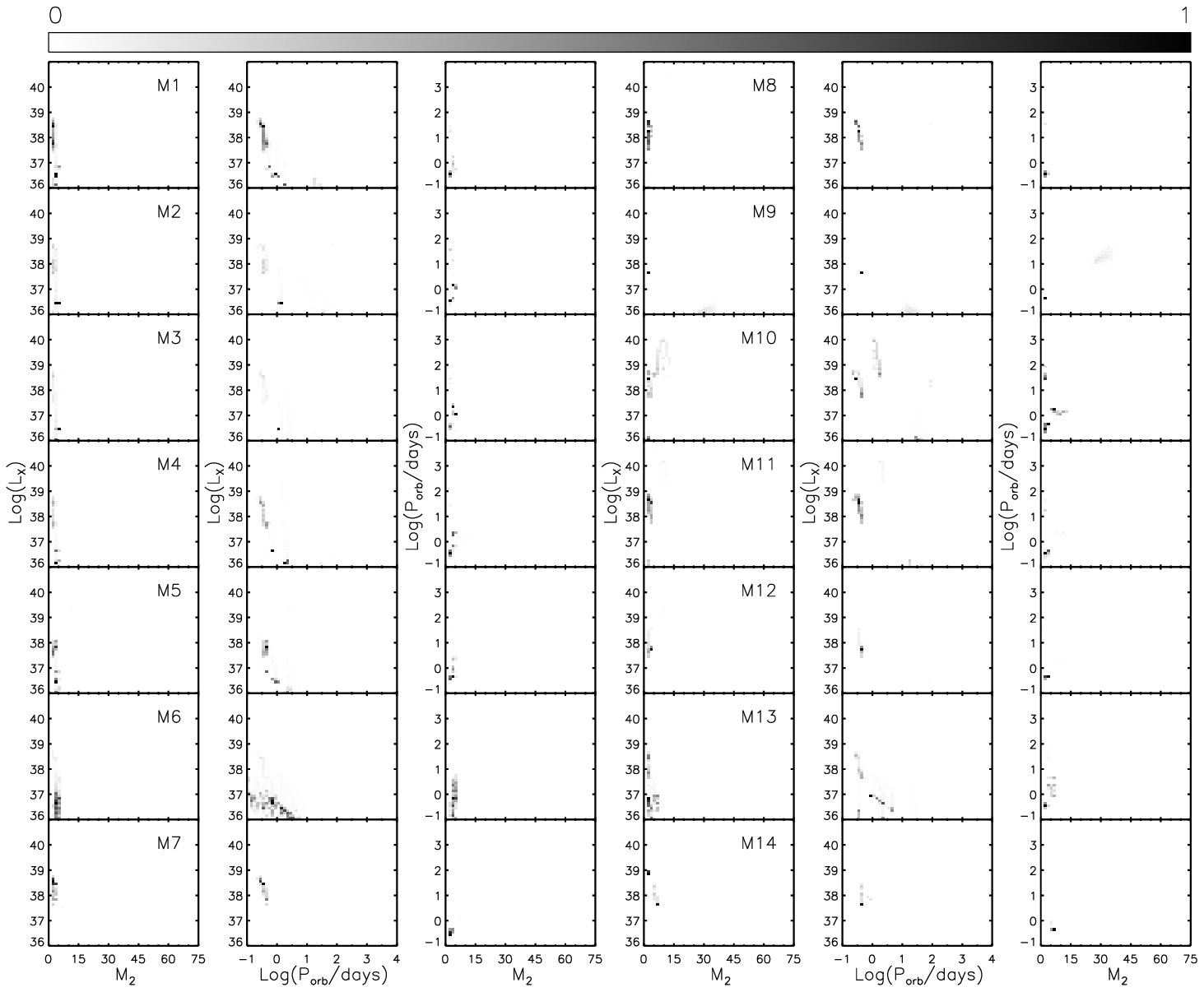}
\caption{Same as in Fig.~5 but for sources in the region of $100<R<1000$ pc
(regions C).}
  \label{Fig. 6}
\end{figure*}

In order to further check the constraints on the value of $\alpha_{\rm CE}$,
we also adopt several models with different assumptions under two typical
values of $\alpha_{\rm CE}$ ($\alpha_{\rm CE}=0.9$ for models M1-M7 and
$\alpha_{\rm CE}=0.2$ for models M8-M14, see Table~2). The results are presented
in Fig.~3 (cumulative distributions, same as in Fig.~1) and Fig.~4 ($L_{\rm X}$  vs. $R$
distributions, same as in Fig.~2), respectively. One can see that models with
$\alpha_{\rm CE}=0.9$ (i.e., models M1-M6 except M7, possibility 
$\sim 10^{-4}$, see Table~4) can match the observation generally while models 
with $\alpha_{\rm CE}=0.2$ (i.e., models M8-M14) still fail. We note that 
models M8-M14 all have extremely small possibilities except models M12 and M13, 
both of which are further ruled out by the $L_{\rm X}$ vs. $R$ distributions.
Stronger stellar wind (i.e., models M7 and M14) leads to too few HMXBs, as well as
ones at large displacements, which is not comparable with observations,  so we
do not give further discussions on it but only present the corresponding informations for reference.
Note that models M1-M6 all have similar $L_{\rm X}$ vs. $R$ correlations which are
constructed by both BH-XRBs and NS-XRBs. It is clear from Fig.~4 that BH-XRBs mainly
dominate at small-offset ($10<R<100$ pc) regions, with luminosities covering a broad
range, from high luminosities ($L_X>10^{38}$ ergs$^{-1}$, named as regions A), to low
luminosities ($10^{36}<L_{\rm X}<10^{38}$ ergs$^{-1}$, named as regions B).
While NS-XRBs spread in much broader spatial regions, as far as about 1 kpc, however
their maximum luminosities are relatively lower than that of BH-XRBs in all of the models,
due to a lower Eddington accretion rate limit, as expected. Similarly we define the regions
with the spatial offset of $100<R<1000$ pc as region C in order for further analysis. We
emphasize here that different regions in the $L_{\rm X}-R$ plane are occupied by
different populations of HMXBs, for example regions A generally BH-XRBs, regions
C NS-XRBs, however regions B are dominated by both NS and BH XRBs.

In order to explore the nature of XRBs in different regions, we also
examined their observational properties, such as current mass
$M_2$ and spectral type of the donor star, orbital period
$P_{\rm orb}$, and system velocity distributions. Figures~5 and 6
show the $L_{\rm X}-M_2$, $L_{\rm X}-P_{\rm orb}$, and
$P_{\rm orb}-M_2$ distributions of XRBs for models M1-M7 (left three
columns, from top to bottom) and M8-M14 (right three columns, from
top to bottom) in the region of $10<R<100$ pc (i.e., regions A and B)
and $100<R<1000$ pc (regions C), respectively. The detailed source
types in regions A, B and C for models M1-M14 are listed in Tables~5-7,
respectively.

\begin{table*}\begin{center}
\caption{The detailed types of sources in region A ($L_{\rm
X}>10^{38}\rm\, erg\,s^{-1}$, $10<R<100$ pc). Here BH per cent represents the percentage of
BH-XRBs in region A, $\frac{\rm N(>10^{38} erg/s)}{\rm N(>10^{36}
erg/s)}$ represents the percentage of high-luminosity ($L_{\rm
X}>10^{38}\rm\, erg\,s^{-1}$) sources in $10<R<100$ pc region.
``BH(NS)MS" and ``BH(NS)HeMS" represent
BH(NS)-XRBs with MS companions and BH(NS)-XRBs with HeMS
companions, respectively. }
\renewcommand\tabcolsep{1pt}
\begin{tabular}{ccccccc}\hline\hline
   Model & BH per cent  & $\frac{\rm N(>10^{38} erg/s)}{\rm N(>10^{36} erg/s)}$ & $\frac{\rm BHMS}{\rm BH}$ & $\frac{\rm BHHeMS}{\rm BH}$
   & $\frac{\rm NSMS}{\rm NS}$ & $\frac{\rm NSHeMS}{\rm NS}$ \\ \hline
      M1 & 67  & 14 & 0  & 99 & 100  & 0\\
      M2 & 37  & 7   & 1  & 99 & 95  & 5\\
      M3 & 90  & 17 & 0  & 99 & 81  & 19\\
      M4 & 67  & 21 & 1  & 99 & 92  & 8\\
      M5 & 61  & 16 & 0  & 99 & 95  & 5\\
      M6 & 39  & 10   & 1  & 99 & 10  & 90\\
      M7 & 0  & 7 & 0  & 0 & 49  & 51\\      \hline
      M8 & 72  & 2   & 0  & 99 & 100  & 0\\
      M9 & 86  & 0   & 100 & 0   & 100    & 0\\
      M10 & 51  & 8   & 0  & 99 & 95  & 5\\
      M11 & 53  & 11   & 0  & 99 & 97  & 3\\
      M12 & 54  & 23 & 1  & 99 & 98  & 2\\
      M13 & 1  & 10  & 1  & 99 & 16  & 84\\
      M14 & 0  & 6 & 0  & 0 & 16  & 84\\\hline
\end{tabular}\label{tab:da14} 
\end{center}
\end{table*}

\begin{table*}\begin{center}
\caption{Same as in Table~5 but for sources in region B
($10^{36}<L_{\rm X}<10^{38}\rm\, erg\,s^{-1}$, $10<R<100$ pc). Here
$\frac{\rm N(10^{36}<L_{\rm X}<10^{38} erg/s)}{\rm N(>10^{36}
erg/s)}$ represents the percentage of low-luminosity
($10^{36}<L_{\rm X}<10^{38}\rm\, erg\,s^{-1}$) sources in $10<R<100$ pc
region.} 
\renewcommand\tabcolsep{1pt}
\begin{tabular}{ccccccc}\hline\hline
   Model & BH per cent  & $\frac{\rm N(10^{36}<L_{\rm X}<10^{38} erg/s)}{\rm N(>10^{36} erg/s)}$ & $\frac{\rm BHMS}{\rm BH}$ & $\frac{\rm BHHeMS}{\rm BH}$
   & $\frac{\rm NSMS}{\rm NS}$ & $\frac{\rm NSHeMS}{\rm NS}$ \\ \hline
      M1 & 42   & 86 & 99  & 0 & 26   & 74\\
      M2 & 50   & 93 & 99  & 1 & 12   & 88\\
      M3 & 32   & 83 & 98  & 2 & 9   & 91\\
      M4 & 22   & 79 & 88  & 12 & 3   & 97\\
      M5 & 3  & 84 & 56  & 44 & 7  & 93\\
      M6 & 3  & 90   &94  & 6 & 1  & 99\\
      M7 & 0  & 93 & 0  & 0 & 49  & 51\\     \hline
      M8 & 91   & 98 & 99  & 1 & 69   & 31\\
      M9 & 91   &100& 100  & 0 & 100   & 0\\
      M10 & 61   & 92 & 64  & 36 & 71   & 29\\
      M11 & 89   & 89 & 99  & 0 & 34   & 66\\
      M12 & 37  & 77 & 87  & 13 & 30  & 70\\
      M13 & 16  & 90  & 100 & 0&2 & 98\\
      M14 & 0  & 94 & 0  & 0 & 23  &77\\\hline

\end{tabular}\label{tab:db14} 
\end{center}
\end{table*}

\begin{figure*}
  \centering
   \includegraphics[width=5.5in]{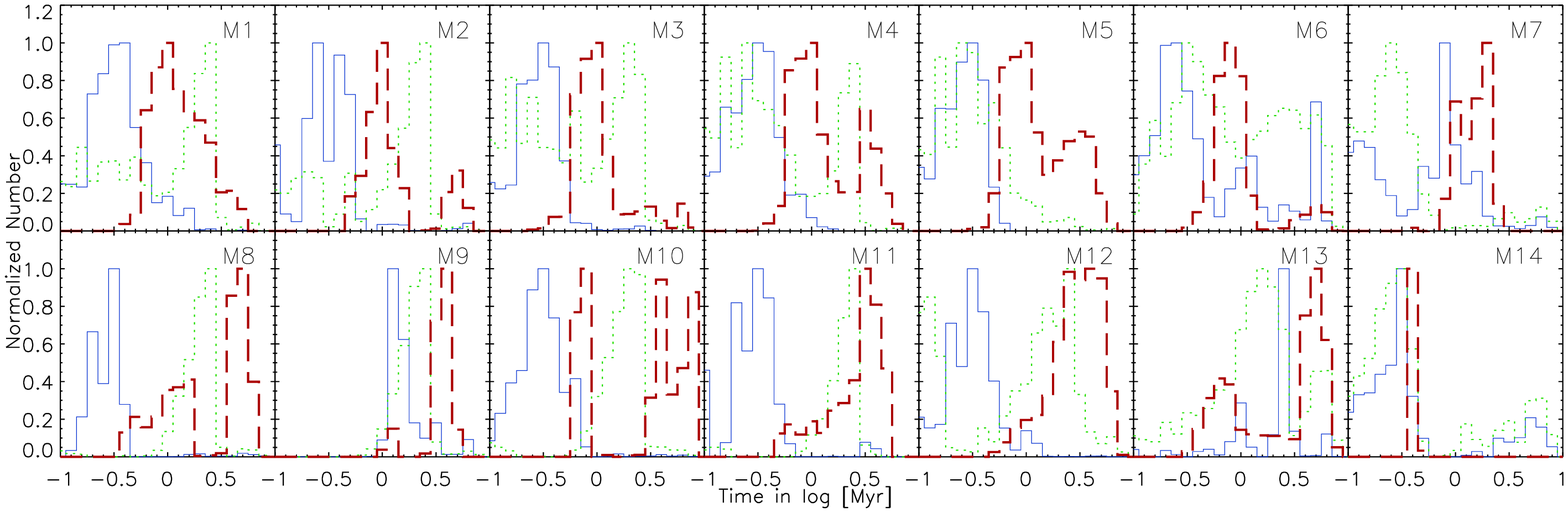}
\caption{The delay time distributions between SN and the turning-on
of X-rays for sources in regions A (solid line), B (dotted
line) and C (thick dashed line), respectively. From left to right are
models M1-M7 (upper panel) and M8-M14 (lower panel),
respectively.}
  \label{Fig. 7}
\end{figure*}

\begin{table*}\begin{center}
\caption{Same as in Table~5 but for sources in region C ($L_{\rm
X}>10^{36}\rm\, erg\,s^{-1}$, $100<R<1000$ pc). $\frac{\rm N(>10^{38}
erg/s)}{\rm N(>10^{36} erg/s)}$ represents the percentage of
high-luminosity ($L_{\rm X}>10^{38}\rm\, erg\,s^{-1}$) sources in
region C.}
\renewcommand\tabcolsep{1pt}
\begin{tabular}{ccccccc}\hline\hline
   Model & BH per cent  & $\frac{\rm N(>10^{38} erg/s)}{\rm N(>10^{36} erg/s)}$ & $\frac{\rm BHMS}{\rm BH}$ & $\frac{\rm BHHeMS}{\rm BH}$
   & $\frac{\rm NSMS}{\rm NS}$ & $\frac{\rm NSHeMS}{\rm NS}$ \\ \hline
      M1 & 6   & 26 & 2  & 98 & 70 & 30\\
      M2 & 7   & 22 & 5  & 95 & 45 & 55\\
      M3 & 6   & 20 & 42& 57 & 28 & 72\\
      M4 & 3   & 25 & 51& 49 & 48 & 52\\
      M5 & 2  & 8 & 30  & 70 & 53  & 47\\
      M6 & 1  & 4   & 4  & 96 & 2  & 98\\
      M7 & 0  & 74 & 0  & 0 & 21  &79\\      \hline
      M8 & 6 & 65 & 5  & 95 & 100 & 0\\
      M9 & 0   & 0   & 0  & 0  &100  & 0\\
      M10 & 61 & 69 &64& 36 & 100 & 0\\
      M11 & 13 & 79 &37& 63 & 100 & 0\\
      M12 & 3  & 14 & 47  & 53 & 91  & 1\\
      M13 & 0 & 5 & 0 & 0 & 13  & 87\\
      M14 & 0  & 64 & 0  & 0 & 0  & 100\\\hline

\end{tabular}\label{tab:dc14} 
\end{center}
\end{table*}

Fig.~5 shows that the XRBs in region A are in short orbital periods (several
hours to $\sim 1$ day), with donors around several solar masses to $\sim
15\,M_{\odot}$ for all models. They are mainly BH-XRBs with helium 
main-sequence (HeMS) donors and NS-XRBs with main-sequence (MS) companions
(see Table~5). The binary velocities are $\sim 100-200\,\rm km\,s^{-1}$ for
all models. Considering that they have relative short evolutionary time-scales
(generally $\sim 0.5$ Myr, see Fig.~7), they can not move too far, even not
farther than 100 pc. The XRBs in region B however show some diversities. They
can be further divided into two subgroups, one with relative short orbital
periods (about days) and less massive ($< \sim 10\,M_{\odot}$) companions,
while the other with donors much more massive ($\sim 30-60\,M_{\odot}$)
and orbits much wider (periods about tens of days to even hundreds of days).
Further analysis indicates that the former group is mainly NS-XRBs, with relative
high speed which peaks at $\sim 200\rm\,km\,s^{-1}$, while the latter one is
mainly low-speed ($< \sim 30\,\rm km\,s^{-1}$) BH-XRBs, which are powered
by stellar winds from massive MS donors, dominating at the low luminosity range
($\sim 10^{36}-10^{37}\,\rm erg\,s^{-1}$).  Considering that they have similar
evolutionary timescales (i.e., of the order of 1 Myr, see Fig.~7), the difference
in the velocities explains the different maximum offsets of NS- and BH-XRBs in
the region. Note that sources in models M8-M11 are dominated by this kind of
low-speed BH-XRBs (see Table~6), which may explain why the majority of sources
in these models are nearby the cluster center, and they can not match the
observations. Fig.~6 shows that, the XRBs in region C all have relatively low-mass
($< \sim 5 M_{\odot}$) companions. Their orbital periods are around several
hours to days for all models. The typical velocities are $\sim 150-300\rm\,km\,s^{-1}$,
larger than those of high-luminosity sources. Considering that they also have
relatively longer evolutionary timescales ($\sim 0.5$ Myr in regions A versus $\sim
1-10$ Myr in regions C, see Fig.~7), both the two aspects determine that they can
move much farther than sources in regions A. We may see from above that, besides
the system velocities at the moment of SN explosion, the spatial offsets of XRBs also depends
on the delay time from the SN to the onset of X-rays. So in Fig.~7 we present the
distribution of the delay time for sources in regions A (solid line), B (dotted line) and
C (thick dashed line), respectively. We normalize each the histograms by the total
number of X-ray sources in each region. It is clear that sources in regions A have the
shortest evolutionary timescales, while the timescales in regions B and C distribute
more broadly. We note sources in regions C may reach the longest evolutionary
timescales, especially for sources in models M8-M13, however they are rare.

\begin{figure*}
  \centering
   \includegraphics[width=4.3in]{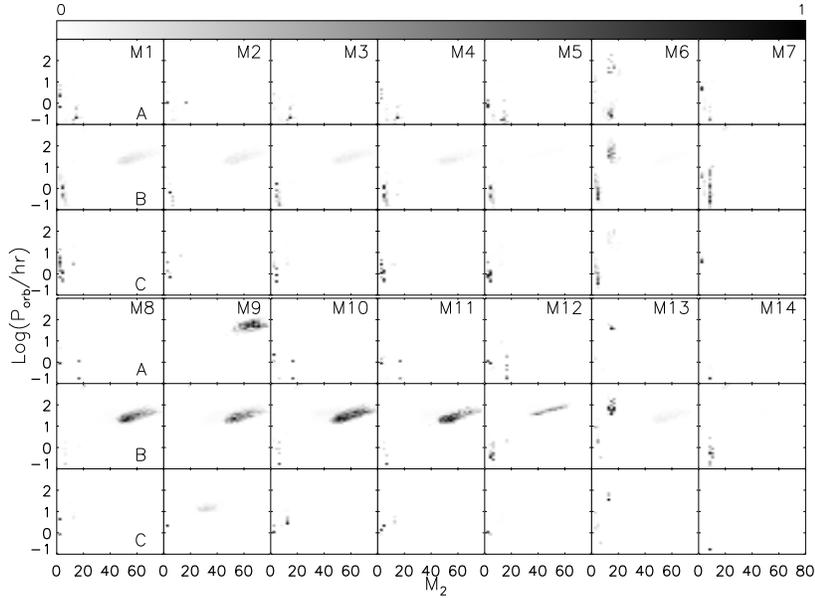}
\caption{The $P_{\rm orb, SNe}-M_{\rm 2, SNe}$ distributions in
regions A, B and C, respectively. From left to right are
models M1-M7 (upper panel) and M8-M14 (lower panel), respectively.}
  \label{Fig. 8}
\end{figure*}

\begin{figure*}
  \centering
   \includegraphics[width=4.3in]{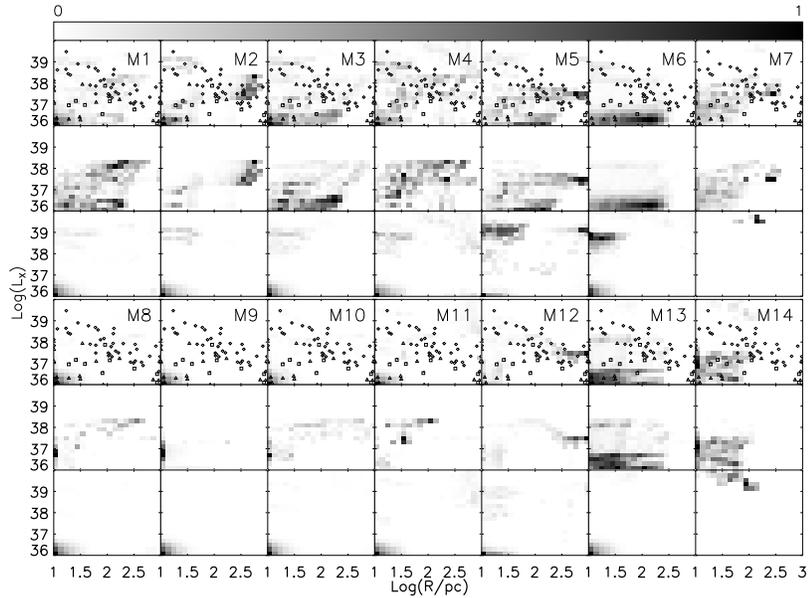}
\caption{Same as in Fig.~4 but with AIC of accreting NSs considered.}
  \label{Fig. 9}
\end{figure*}

We note that the differences of binary velocities among models are still tightly related
to the companion masses $M_{\rm 2, SNe}$, and the orbital periods $P_{\rm orb, SNe}$
(or orbital velocity) at the moment of SN explosions (see Eq.~5), as already stated in
\citet{zuo10}. The CE parameter $\alpha_{\rm CE}$, as a key parameter that affects
the binary orbit, can not only determine the types of XRB populations, but also affect
the orbital velocity immediately after the CE, hence the global velocity of the binary
system. These phenomena are demonstrated clearly in Fig.~8, i.e., the $P_{\rm orb, SNe}
-M_{\rm 2,SNe}$ distribution in regions A, B and C for models M1-M7 (upper panel) and
M8-M14 (lower panel), respectively.

It is clear that short period pre-SN systems in models M8-M13 are much less than in models
M1-M6, resulting in much less high speed sources. It is because of the fact that larger
values of $\alpha_{\rm CE}$ (i.e., models M1-M6) can prevent coalescence  of a NS/BH
and the companion in a compact binary during the unstable mass transfer processes,
in favor of the formation of tight XRBs, hence more sources with larger orbital and system
velocities. While it is not the case for models M8-M13 where BH-MS XRBs dominate generally.
They mainly have longer-period (hence smaller orbital velocity) and more massive companions,
resulting in smaller system velocity (hence smaller offset from the parent cluster)
than others. These facts clearly imply that the CE parameter $\alpha_{\rm CE}$,
by affecting the binary orbit, plays an important role in the kinematic motion and
spatial distributions of XRBs, demonstrating itself as distinct $L_{\rm X}$ vs. $R$
distributions. And conversely, the well measured $L_{\rm X}$ vs. $R$ distribution
can make a good decision for the precise choice of the CE parameter. Our work
motivates further efforts to explore the spatial distributions of XRBs, as well as
their source types in nearby star-forming galaxies.

We note that in the above cases, we did not consider BHs formed from
accretion-induced  collapse (AIC) of NS systems, however whether AIC
of NSs happens or not is still in controversy. In order to examine this effect,
we also consider the NS$\rightarrow$BH AIC formation channel. The results
are presented in Fig.~9, which is the same as in Fig.~4, except AIC of NSs
considered here. One can see that, there are some differences
when considering NS AICs. The most remarkable feature is that several
BH XRBs appear in region C, especially in model M4, too many ultra-luminous
X-ray sources (ULXs) appear in this region, which destroys the correlation badly.
They all have relatively high speeds, typically $\sim 150-300\rm\,km\,s^{-1}$.
Additionally in the case of  $\alpha_{\rm CE}=0.2$ (i.e., models M8-M14),
the $L_{\rm X}$ vs. $R$ distribution  still can not be well constructed.
We suggest that precise velocity measurement of BH-XRBs are promising
to discriminate different BH formation channels.

\begin{figure*}
  \centering
   \includegraphics[width=4.3in]{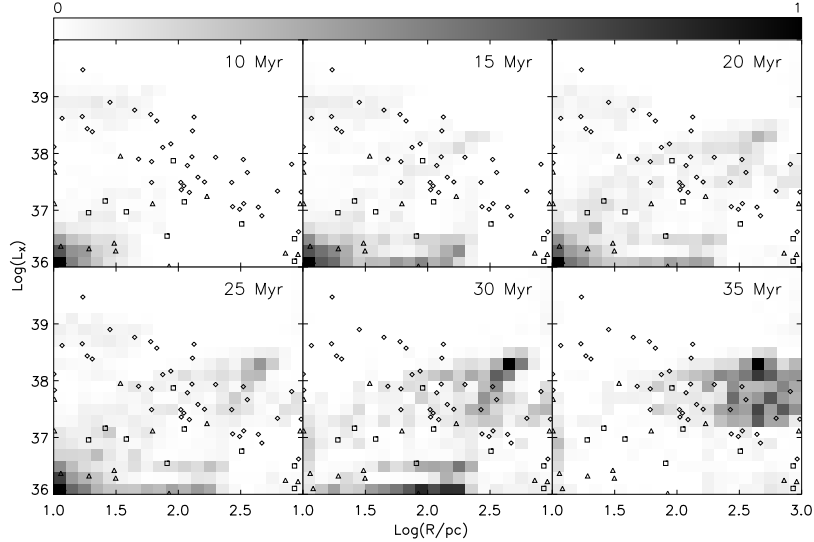}
\caption{The evolution of the $L_{\rm X}-R$ distribution for model M1. The time is
set at 10, 15, 20, 25, 30 and 35 Myr since the beginning of star formation, respectively.}
  \label{Fig. 10}
\end{figure*}

The evolution of $L_{\rm X}$ vs. $R$ for model M1 is presented in Fig.~10. It
can be seen that as the star formation proceeds, the $L_{\rm X}$ vs. $R$
correlation is gradually built up till the time of about 20 Myr, then since the
star formation quenches, the correlation disappears gradually as time goes 
on. So the $L_{\rm X}$ vs. $R$ correlation cannot hold all the time for
individual cluster. In addition, we also examined the cases with/without 
Be-XRBs in the populations, and found that the impact of Be-XRB is minor over 
the whole evolution interested, due to its short transient characteristics.

Our modeled $L_{\rm X}$ vs. $R$ correlation
and its constraints on $\alpha_{\rm CE}$ is promising and sound, but
they are still subject to some uncertainties and simplified
treatments. Firstly, the binaries we evolved are all primordial ones,
however they may also be produced by dynamical interactions,
especially in dense star clusters. However in this formation channel,
the massive stellar BHs which enter RLOF, thought as possible ULXs
may have much longer evolution timescales \citep[about several tens of
Myr to $\sim 100$ Myr since star formation,][and see references therein]
{Mapelli13}, which is not likely the cases we studied here.
Another simplified treatment is taken for the natal kicks. The role of
natal kicks is essential for our studies, however little is yet known about either the
amount of natal kicks or its distributions, especially for BH kicks
\citep[][and references therein]{bp95,bps95,gcpp05,fragos09}.
It is reasonable to assume that BH receives a smaller natal
kick in the absence of SN than in a SN scenario \citep{fk01,l10}.
However recent work by \citet{rds12} indicates that, in order
to explain the observed distribution of LMXBs with BHs, natal
kicks of BHs seem to be similar to that of NSs. The role of natal
kicks, especially the amount and distribution of BH kicks
deserves a thorough investigation, which will be presented in the
forthcoming paper. Despite all of this,
we note that our results are generally consistent with
previous studies concerning the CE evolution. For example,
a high value of  $\alpha_{\rm CE}$ is required in order to
account for the observed WDMS PCEBs \citep{dkw10}, the shape
of the delay-time distribution and the birth rate of SNe Ia for
the double degenerate systems \citep{my12}, and HMXB
XLFs \citep{zuo13b}, etc. Incorporating the kinematic evolution of
HMXBs, we can give a more feasible-to-check observational
properties of HMXBs, which may help understand the CE
evolution and to constrain the value of $\alpha_{\rm CE}$.

\section{SUMMARY}

We have used a BPS code to model the luminosity versus
displacement correlation of HMXBs in star-burst galaxies. We
used the apparent correlation to constrain the CE parameter
$\alpha_{\rm CE}$, and find that within the framework 
of the standard energy formula for CE and core definition at mass 
$X=10$\%, a high value of $\alpha_{\rm CE}$ around $0.8-1.0$ 
is more preferable while $\alpha_{\rm CE}< \sim 0.4$ can not 
match the observation. We caution that alternative definitions for the core may 
change the value of $\lambda$ by a factor of about 
two order of magnitude (for example, `entropy profile' 
definition), which may translate directly the same uncertainty 
to the value of $\alpha_{\rm CE}$. We split the $L_{\rm X}$ 
vs. $R$ plane into three regions, named as regions A, B and C, 
specify the detailed properties of HMXB populations in each 
region. The results are listed in Tables~5-7, and summarized below.

(1) The high-luminosity ($L_X>10^{38}$ erg\,s$^{-1}$) sources in
the $10<R<100$ pc region are mainly BH-HeMS and NS-MS XRBs.
They all have short orbital periods (several hours to $\sim 1$ day).
Their donor masses are around several solar masses to $\sim 15\,M_{\odot}$.
The system velocity is $\sim 100-200\rm\,km\,s^{-1}$.

(2) The low-luminosity ($10^{36}<L_X<10^{38}$ erg\,s$^{-1}$) sources in
the $10<R<100$ pc regions contain two species, i.e., high
speed (peaks at $\sim 200\rm \,km\,s^{-1}$) NS-XRBs with
less massive ($< \sim 10 \,M_{\odot}$) companions in short
orbital periods (about days), and low speed ($< \sim 30 \,\rm km\,s^{-1}$)
BH-XRBs powered by stellar winds from massive ($\sim 30-60\, M_{\odot}$)
MS donors in wide orbits (periods about tens of days to even hundreds of days).

(3) The XRBs in the $100<R<1000$ pc regions are mainly NS-XRBs.
They mainly have relatively low-mass ($< \sim 5\, M_{\odot}$)
companions.. Their orbital periods are about several hours to
days, and their system velocities are $\sim$ 150-300 km\,s$^{-1}$.

Our studies show clearly that the CE parameter $\alpha_{\rm CE}$,
by governing the binary orbit during the CE, affects not only the outcome
of the population, but also the kinematic motion, hence the spatial distribution of XRBs,
revealing as distinct $L_{\rm X}$ vs. $R$ distributions.
So better measurements of $L_{\rm X}$ vs. $R$ distribution, as well as
the detailed properties of sources are promising to give a better constraint on
the CE efficiency parameter. Our work motivates further high-resolution
optical and X-ray observations of HMXB populations in nearby star-forming galaxies.

\section*{Acknowledgements}
This work was supported by the National Natural Science
Foundation (grants 11103014 and 11003005), the Research Fund for
the Doctoral Program of Higher Education of China (under grant number 20110201120034),
the Natural Science Foundation of China (under grant number 10873008), the National
Basic Research Program of China (973 Program 2009CB824800), the Fundamental Research Funds for the
Central Universities and National High Performance Computing Center (Xi'an).

\label{lastpage}
\end{document}